**Title:**
Markov blankets in the brain

**Authors:**
Inês Hipólito[1,2,3,6]
Maxwell J. D. Ramstead[4,5,6]
Laura Convertino[6,7]
Anjali Bhat[6]
Karl Friston[6]
Thomas Parr[6]

**Affiliations:**
1. Centre for Neuroimaging Sciences, Department of Neuroimaging, King's College London, United Kingdom
2. Mind, Brain Imaging and Neuroethics, Institute of Mental Health Research, University of Ottawa, Ottawa, Canada.
3. Department of Philosophy, University of Wollongong, Wollongong, Australia
4. Division of Social and Transcultural Psychiatry, Department of Psychiatry, McGill University, Montreal, Quebec, Canada.
5. Culture, Mind, and Brain Program, McGill University, Montreal, Quebec, Canada.
6. Wellcome Centre for Human Neuroimaging, University College London, London, United Kingdom.
7. Institute of Cognitive Neuroscience (ICN), University College London, London, United Kingdom.

**Abstract**
Recent characterisations of self-organising systems depend upon the presence of a 'Markov blanket': a statistical boundary that mediates the interactions between what is inside of and outside of a system. We leverage this idea to provide an analysis of partitions in neuronal systems. This is applicable to brain architectures at multiple scales, enabling partitions into single neurons, brain regions, and brain-wide networks. This treatment is based upon the canonical micro-circuitry used in empirical studies of effective connectivity, so as to speak directly to practical applications. This depends upon the dynamic coupling between functional units, whose form recapitulates that of a Markov blanket at each level. The nuance afforded by partitioning neural systems in this way highlights certain limitations of 'modular' perspectives of brain function that only consider a single level of description.

# Introduction

Scientific investigation in neurobiology often begins – perhaps only implicitly – by partitioning the brain into functional units. While the most obvious partition of neural tissue is into individual neurons, the same process takes place over a range of spatiotemporal scales. The division of the cortical surface into Brodmann areas represents one such carving up of neural tissue (Brodmann, 2007; (Zilles and Amunts 2010). Brodmann maps have enduring practical implications. For example, the Talairach Atlas (Talairach and Szikla, 1980), commonly in use in neuroimaging, may be seen as a direct descendent. In this setting, the assumption is that brain function depends upon interactions between architectonically defined brain regions (Lazar, 2008). This assumption underwrites the study of connectivity in the brain, as we need to know what is being connected. Effective connectivity studies go as far as to distinguish between connections 'intrinsic' or 'extrinsic' to a given region (or cortical column) (Tsvetanov, Henson et al. 2016, Zhou, Zeidman et al. 2018). Again, this rests upon drawing boundaries around parts of a brain. Our focus in this paper is on how such boundaries may be licensed.

A prominent justification for drawing boundaries – from the last century – is the 'modularity of mind' paradigm (Fodor 1983), which itself inherits from the phrenology of the century before that (Gall and Lewis 1835). This conceptualisation of cognitive processes depends upon discrete cognitive units that interact with one-another, which might manifest in the tissue engaged in cognitive operations. An important limitation of this paradigm is that it typically only considers a single level of description, neglecting the rich intrinsic and extrinsic dynamics across regions and microcircuits. In addition, the philosophical assumptions of modular perspectives on neuronal organisation have been criticised (Friston 2002; Colombo, 2013; Palecek, 2017; George and Sunny, 2019; Hipólito and Kirchhoff, 2019). In short, this calls for a more nuanced treatment of partitions and functional interactions.

A growing literature leverages the *Markov blanket* construct (Pearl 1988) to formalise dynamic coupling in physical and biological systems (Friston, 2019, Hipólito, 2019, Ramstead et al., 2018a, Ramstead et al., 2019a, Palacios et al., 2020, Kirchhoff et al., 2018). This construct is a description of the dependencies within and between random dynamical systems – like the brain – that sets a boundary between the inside and outside of each system. Here, we focus upon the Markov blankets implicit in the kinds of models used practically in investigating brain function. Specifically, we examine the dynamics implied by neural mass models[1] of the kind that underwrite Dynamic Causal Modelling (DCM) (Bastos, Usrey et al. 2012, Moran, Pinotsis et al. 2013). Building from this to the connectivity of a canonical cortical microcircuit, we set out a series of Markov blanketed structures at increasing spatial scales.

This approach endorses the segregation of the brain into regions but also emphasises the absence of a privileged scale of description at which 'modules' might be defined. By selecting a Markov blanket, we implicitly identify the variables that define the simplest element of our system at a given scale. It follows that, depending on the scale of interest, the variables comprising the Markov blanket will be different. For a single neuron, the blanket includes the presynaptic and postsynaptic membrane potentials that mediate its interactions with other neurons. For cortical columns, the blanket will include neural populations mediating interactions between different columns. In principle, the identification of functional boundaries can proceed at finer (ion channels and molecules) and coarser (networks, brains, and people) levels.

While identifying blankets at each level may seem an abstract exercise, it has important implications for experimental neuroscience. Specifically, it offers an important part of the conceptual analysis we need to ensure our hypotheses make sense (Nachev and Hacker 2014). For example, if we want to know whether condition specific differences in measured brain activity are mediated by changes in 'intrinsic' or 'extrinsic' connectivity (Zhou, Zeidman et al. 2018), we need to be able to define what we mean by these terms, and to say what they are intrinsic or extrinsic to. We aim to make this explicit in a series of examples.

The aim of this paper is to argue that an appeal to the Markov blanket construct provides a formal basis for partitioning the brain into functional units – from individual neurons to functional assemblies of neurons, through to independent brain regions and networks of regions. In particular, we will argue that a recursively iterated version of the formalism, where each component of a Markov blanketed system is itself a Markov blanketed system, is apt for the task. This paper comprises four parts. The first provides a brief overview of the Markov blanket construct and its relevance to a dynamical setting. The second section zooms in on the individual neurons and illustrates how synaptic dynamics conform to the conditional independence structure of a Markov blanket. The third takes a more detailed look at the asymmetries of the neuronal Markov blanket, and emphasises the need for these to be replicated at the network level. The fourth section zooms out and shows how the same structure is recapitulated at larger spatial scales.

---

[1] We will occasionally appeal to technical terms that are in common usage in this field. Please see the glossary for definitions.

## 1. Markov blankets

The Markov blanket construct, which underwrites the current proposal, was introduced into the literature by Pearl (1988) in the context of statistical inference. To distinguish a set of systemic (or internal) states from their embedding environment (of external states), a third set of states are implied. These are the Markov blanket (Friston, 2013). The Markov blanket consists of sensory states, which affect but are not affected by internal states; and active states, which affect but are not affected by external states (Figure 1). This implements a form of conditional independence between internal states and external ones.

A Markov blanket ($b$) around internal states $\mu$ – where all other (external) variables are labelled $\eta$ – is defined as the set of variables that renders $\mu$ conditionally independent from $\eta$. Mathematically, this is written as follows:

$$\mu \perp \eta \mid b \Leftrightarrow p(\mu, \eta \mid b) = p(\mu \mid b) p(\eta \mid b) \qquad (1)$$

Equation 1 illustrates this dependency structure in the factorisation of the joint distribution conditioned on blanket states into two independent distributions; by definition, two variables are conditionally independent if and only if their joint probability, conditioned on some third variable, is equal to the product of their individual probability conditioned on that third variable. It is common to speak of the random variables separated in this way by Markov blankets – and the associated conditional dependencies – in terms of 'parents' and their 'children', where 'parent' nodes cause their children. A Markov blanket is then the set of the parents, the children, and the parents of the children of the variable in question. An alternative way to frame this is to think of the parents as mediating the influence of external states on internal states (i.e., sensory states) and the children (and their parents) as mediating the influence of internal states on external states (i.e., active states). This suggests a separation of blanket states into active ($a$) and sensory ($s$) states.

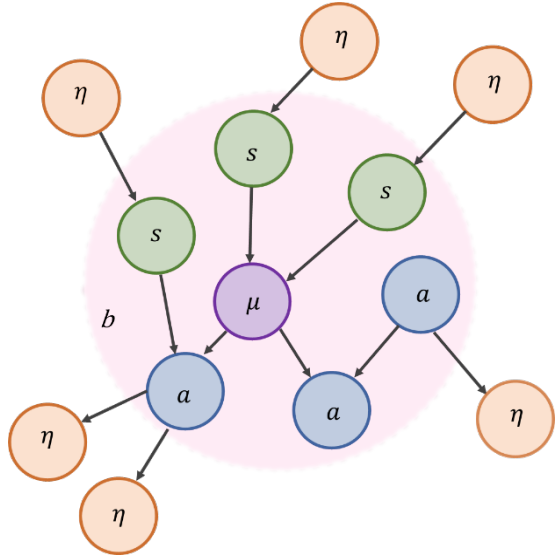

**Figure 1**. Markov blanket. A Markov blanket highlights open systems exchanging matter, energy or information with their surroundings. Variables *η* are conditionally independent of variables *μ* by virtue of its Markov blanket (*b*). If there is no route between two variables, and they share parents, they are conditionally independent. Arrows go from parents to children. We will use the colour-scheme in this figure consistently through subsequent figures.

In a dynamical setting (Friston, Da Costa et al. 2020), Equation 1 means that the average (represented in bold) rate of change of each component of a Markov blanketed system can only depend on two other sorts of state in order to preserve the structure of Equation 1. This is shown in Equation 2 and Figure 2:

$$\begin{aligned}\dot{\boldsymbol{\mu}} &= f_\mu(\boldsymbol{\mu},\mathbf{s},\mathbf{a}) \\ \dot{\mathbf{a}} &= f_a(\boldsymbol{\mu},\mathbf{s},\mathbf{a}) \\ \dot{\boldsymbol{\eta}} &= f_\eta(\boldsymbol{\eta},\mathbf{s},\mathbf{a}) \\ \dot{\mathbf{s}} &= f_s(\boldsymbol{\eta},\mathbf{s},\mathbf{a})\end{aligned} \qquad (2)$$

Equation 2 means that a dynamical system that preserves the conditional dependency structure of the Markov blanket ensures that the flow of internal and external states do not depend upon one another; i.e., that internal states cannot influence sensory states, and that external states cannot influence active states. Additionally, note that the Markov blanket structure is preserved if dependencies are lost (e.g., if the active states were not influenced by sensory states), but not if they are gained, since that would undo the conditional independence. We will see over the next few sections that this structure can be identified at numerous levels of a neural mass model (David and Friston 2003; Pinotsis et al. 2014; Moran et al. 2016).

Before we move on, it is worth briefly unpacking the reason for the names of the variables. While the Markov blanket formulation in general and applies to any random variables, recent work has leveraged Markov blankets to talk about the structure of exchanges between an organism and its environment (Friston, 2013; Kirchhoff et al. 2018; Parr and Friston, 2018,) and to describe self-organisation across spatial and temporal scales (Hipolito, 2019, Ramstead et al., 2018b, Palacios et al., 2017). In this context, we associate the variable of interest with the internal states of a Markov blanket; which allows us to think of the 'parents' of that variable as mediating the influence of external states on internal states (i.e., as sensory states) and of its 'children' and the 'parents of the children' as mediating the influence of internal states on external states (i.e., as active states). This conception of the Markov blanket as the mediating influence of external states on internal states through the effects of sensory and active states resonates with the action-perception cycles typically considered in cognitive systems (Fuster, 1990; Parr and Friston, 2017; Parr and Friston, 2018). This is the reason for the words 'active' and 'sensory'. While it may seem strange to use these terms for interactions at cellular or network levels, it should be emphasised that these are simply names for statistical constructs.

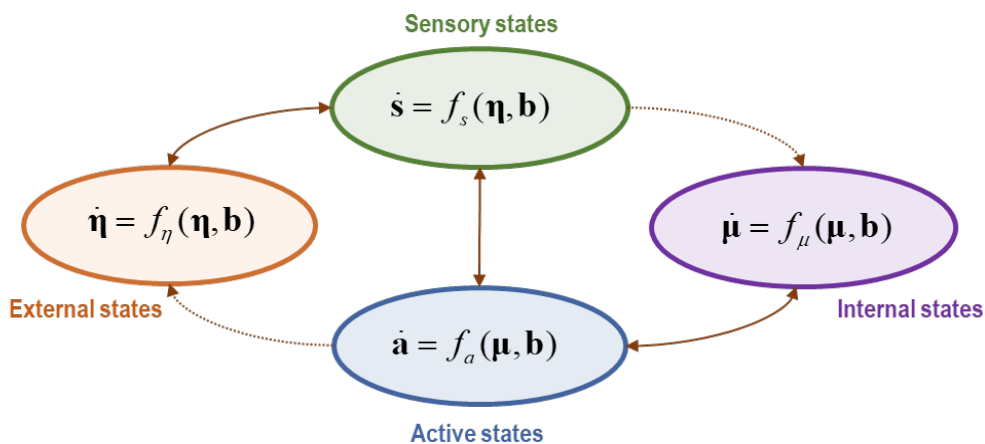

**Figure 2.** This schematic illustrates the partition of states into internal states (purple) and hidden or external states (orange) that are separated by a Markov blanket – comprising sensory (green) and active states (blue). Specifically, it focuses on the dynamical formulation of Equation 2. Directed influences are highlighted with dotted connectors. Autonomous states are those states that are not influenced by external states, while particular states constitute a particle; namely, autonomous and sensory states – or blanket and internal states. Sensory states,

active states and internal states comprise the particular states that are constitutive of a functional neuronal unit (for more detail see Hipólito 2019).

## 2. Neurons and their Markov blankets

In this section, we consider the partition of brain tissue into neurons. From a dynamical perspective, this means finding equations of motion consistent with Equation 2 and Figure 2. We know that synaptic dynamics conform to the conditional independence structure of a Markov blanket, as the internal states (e.g., conductance of ion channels) of one neuron are distinguishable from the same states of other neurons, but interact through presynaptic and postsynaptic voltages. The implied partitioning of tissue into Markov blanketed neurons allows neurons to change their behaviour without losing their identity.

Figure 3 shows explicitly how synaptic dynamics conform to the conditional independence structure of a Markov blanket. This is based upon the neural dynamics formalised in dynamic causal models (Bastos, Usrey et al. 2012, Moran, Pinotsis et al. 2013). This is one of many models of neural dynamics, and we have summarised common alternatives in Table 1 with varying degrees of biophysical detail. As we said earlier, the existence of a Markov blanket implies a partition of states into external, sensory, active and internal states. The dynamics set out in Figure 3 assign these labels to the variables that preserve the form of Equation 2 – i.e., internal states evolve based upon internal and blanket states but not external states, active states do not depend upon external states, and so on.

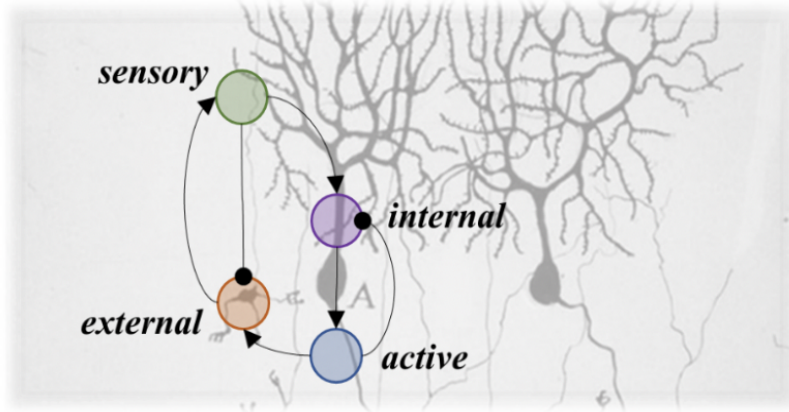

post-synaptic conductance $\quad \dot{\mu} = \frac{1}{\tau}(A_\mu \sigma(s) - 2\mu - \frac{1}{\tau}a)$

post-synaptic voltage $\quad \dot{a} = \mu$

pre-synaptic conductance $\quad \dot{\eta} = \frac{1}{\tau}(A_\eta \sigma(a) - 2\eta - \frac{1}{\tau}s)$

pre-synaptic voltage $\quad \dot{s} = \eta$

**Figure 3** *Neuronal Markov blankets*. This figure illustrates a Markov blanket separating the membrane conductances of a pair of neurons (or between one postsynaptic neuron and all presynaptic neurons). The **A** terms here are constants that act as connectivity strengths from the active state of one neuron to the external state of another (**A**$_\eta$), and from the sensory states of the latter to the internal states of the former (**A**$_\mu$). When many neurons are in play, this becomes a connectivity matrix. The σ-function is a sigmoid shape and may be thought of as converting potentials to firing rates. An interesting feature of this structure is that the sensory states, from the perspective of a given neuron, can arise from many different external states (other neurons) while the active states (membrane depolarisation) depend only on the conductance (internal state) of the neuron being depolarised. Normal arrowheads indicate an excitatory influence, while round arrowheads show inhibitory influences.

It is worth noting that Markov blankets do not trivially correspond to the boundaries of the neuronal cells. Rather, the idea is that the presence of a Markov blanket ensures the influences of blanket variables (here, membrane potentials) vicariously enable internal and external states (ion channel conductance) to communicate. This is fundamental because it means that internal and external states, though not influencing each other directly, are the common units that, when coupled, will determine the large-scale network. Moreover, as the blanket is defined in terms of dynamics as opposed to physical boundaries, which would correspond to the cell membrane at the neuronal level, we start to see how the same formalism applies even in the absence of clear spatial boundaries (Friston, 2013; Kirchhoff et al. 2018). At the neuronal level of description, the Markovian demarcation is not insulation of internal states, but rather a way of highlighting (statistically) which states are relevant for the current investigation (Friston 2019; Hipólito 2019). Ultimately, the dependencies induced by Markov blankets create a circular causality: external states, such as the presynaptic conductance, cause changes in internal states, such as the postsynaptic conductance, via sensory states, i.e. presynaptic voltage, while the internal states couple back to the external states through active states, i.e. the postsynaptic voltage.

**Table 1 – Neural models and their blankets**

| Model | Dynamics | States | Citation |
|---|---|---|---|
| Hodgkin–Huxley | $\dot{\mathbf{a}} = \frac{1}{C}\left(\mathbf{s} - \bar{g}\boldsymbol{\mu}^n \cdot (\mathbf{a} - v)\right)$ <br> $\dot{\boldsymbol{\mu}} = \alpha(\mathbf{a})(1-\boldsymbol{\mu}) - \beta(\mathbf{a})\boldsymbol{\mu}$ <br> $\dot{\mathbf{s}} = f_s(\boldsymbol{\eta})$ <br> $\dot{\boldsymbol{\eta}} = f_\eta(\mathbf{a}, \mathbf{s}, \boldsymbol{\eta})$ | $\mathbf{a}$ – Membrane potential <br> $\boldsymbol{\mu}$ – Ion channels <br> $\mathbf{s}$ – Injected current <br> $\boldsymbol{\eta}$ – Electrophysiologist | (Hodgkin and Huxley 1952) |
| FitzHugh–Nagumo | $\dot{\mathbf{a}} = \mathbf{a} - \frac{1}{3}\mathbf{a}^3 - \boldsymbol{\mu} + \mathbf{s}$ <br> $\dot{\boldsymbol{\mu}} = \frac{1}{\tau}(\mathbf{a} + \alpha - \beta\boldsymbol{\mu})$ <br> $\dot{\mathbf{s}} = f_s(\boldsymbol{\eta})$ <br> $\dot{\boldsymbol{\eta}} = f_\eta(\mathbf{a}, \mathbf{s}, \boldsymbol{\eta})$ | $\mathbf{a}$ – Membrane potential <br> $\boldsymbol{\mu}$ – 'Recovery' variable <br> $\mathbf{s}$ – Injected current <br> $\boldsymbol{\eta}$ – Electrophysiologist | (FitzHugh 1955, Nagumo, Arimoto et al. 1962) |
| Morris–Lecar | $\dot{\mathbf{a}} = \frac{1}{C}\left(\mathbf{s} - \bar{g}\boldsymbol{\mu} \cdot (\mathbf{a} - v)\right)$ <br> $\dot{\boldsymbol{\mu}} = \frac{1}{2\tau(\mathbf{a})}\left(1 + \tanh\left(\frac{1}{u}(\mathbf{a} - v)\right) - 2\boldsymbol{\mu}\right)$ <br> $\dot{\mathbf{s}} = f_s(\boldsymbol{\eta})$ <br> $\dot{\boldsymbol{\eta}} = f_\eta(\mathbf{a}, \mathbf{s}, \boldsymbol{\eta})$ | $\mathbf{a}$ – Membrane potential <br> $\boldsymbol{\mu}$ – Potassium channels <br> $\mathbf{s}$ – Injected current <br> $\boldsymbol{\eta}$ – Electrophysiologist | (Morris and Lecar 1981) |

## 3. Blanket asymmetries

This section deals with the way in which neurons – the basic units of Section 2 – may be connected together to form microcircuits (David and Friston 2003; Moran et al. 2013; Friston et al. 2019; Coombes and Byrne 2019) – which form the basic unit of Section 4. Specifically, we emphasise the key role of asymmetric interactions between blanketed structures. First, we take a step back to briefly highlight the way in which neurons may be studied in isolation. Neurons – as complex, dynamic systems – are highly sensitive to initial conditions, exhibiting organised patterns that result from localised interactions without centralised control (shown schematically in Figure 4). These non-linear interactions may be studied through electrophysiological experiments on single neurons. Typically, this means using voltage clamp experiments and injecting electrical currents. A few examples of physiologically detailed models to account for these non-linear interactions are outlined in Table 1, and include the Hodgkin-Huxley model. This has many moving parts and is therefore rarely used in studies of connected neural populations – where dynamics more akin to those in Figure 3 predominate – but is a good starting point in understanding how sensory states influence the internal state dynamics. This will be essential when we move to sensory states generated by other neural populations in a network.

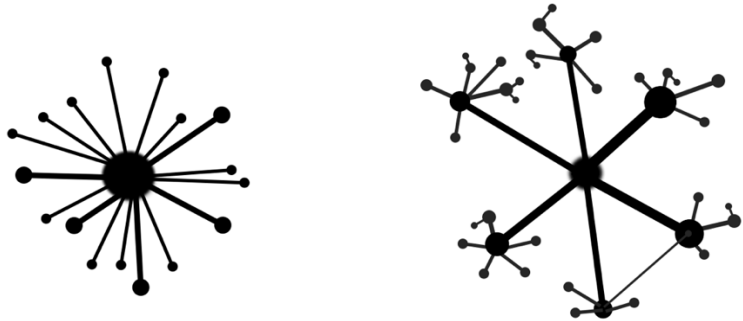

**Figure 4**. On the left a centralised system, representing a central controller. This model is typically motivating modular theories, where a central cognitive system is bounded by lower-level modules. On the right, a decentralised system where elements are not isolated from their environment, and the focus is on the dynamics of the relations among properties and elements.

Intuitively, the Hodgkin–Huxley model expresses the evolution of the membrane potential under time-dependent input currents in terms of the equivalent electric circuit[2], with a potential that evolves based upon membrane capacitance and currents. More specifically, the Hodgkin–Huxley model describes how action potentials in neurons are initiated and propagated through a set of non-linear differential equations that approximates the electrical characteristics of excitable neurons in a continuous-time dynamical system (Douglas and Martin 1991). Formulating the Hodgkin-Huxley (and other models) in terms of the constituents of the Markov blankets inherent in voltage-clamp experiments allows us to highlight the specifics of the influence of the external states (electrophysiologists) via sensory states (injected current) on internal states (ion channels), themselves influencing active states (membrane potential). Unpacking the Equation in Table 1 in terms of the specific ion channels, this is:

---

[2] Note that this is not what is meant by the term 'microcircuit', which refers to the 'wiring' of a population of neurons into a local network.

$$\dot{\mathbf{a}} = \tfrac{1}{C}\left(\mathbf{s} - g_K \boldsymbol{\mu}_K^4 (\mathbf{a} - v_K) - g_{Na} \boldsymbol{\mu}_{Na}^3 (\mathbf{a} - v_{Na}) - g_l \boldsymbol{\mu}_l (\mathbf{a} - v_l)\right)$$

$$\dot{\boldsymbol{\mu}}_i = \alpha(\mathbf{a})(1 - \boldsymbol{\mu}_i) - \beta(\mathbf{a}) \boldsymbol{\mu}_i, \, i = (Na, K, l)$$

$$\dot{\mathbf{s}} = f_s(\boldsymbol{\eta})$$

$$\dot{\boldsymbol{\eta}} = f_\eta(\mathbf{a}, \mathbf{s}, \boldsymbol{\eta}) \tag{3}$$

Here, the capacitance ($C$) mediates the influence of an injected current ($\mathbf{s}$) and ion channel currents on the membrane potential ($\mathbf{a}$). This depends upon the ion channels of the system – i.e., the conductance of the sodium (*Na*), potassium (*K*), and leakage (*l*) channels. These depend upon constants (*g*) and the associated internal states ($\boldsymbol{\mu}$). In addition, it depends on the 'reversal' potentials for each channel (*v*) which specify the potentials at which the direction of ionic flow reverses. The internal states for each channel evolve based upon the (functions – *α* and *β* – of the) potential, as voltage-gated channels open and close to increase or decrease the magnitude of this flow. The nonlinearity inherent in Equation 3 facilitates many interesting biophysical phenomena, including bifurcations and limit cycles (Wang, Chen et al. 2007). However, the purpose of this section is to move us away from the single neuron, and towards the kinds of dynamics exhibited by populations of connected neurons. This requires that we consider the blanket states that mediate these connections. The first step is to notice that the sensory state for the single neuron described by the Hodgkin-Huxley model is an experimental intervention, driven by an experimenter (i.e., an electrophysiologist, $\boldsymbol{\eta}$) who injects current and measures the resulting potential. We need to move to a situation where this comes from other neurons. This is afforded simply by the equations of motion set out in Figure 2 for a pair of neurons.

To understand the way in which blankets connect to one another, it is useful to consider that the membrane potential (active state) of a given neuron can only be directly influenced by the conductance (internal states) of that neuron. In contrast, the presynaptic potentials (sensory states) of many other neurons contribute to the internal states. This asymmetry in the blanket states recapitulates that seen in physical systems. Specifically, the position of many different particles (sensory states) can influence the momentum (internal state) of a single particle. However, the position of the particle in question (active state) is only influenced by the momentum of that same particle. This suggests a clear analogy between Newtonian mechanics and neuronal mechanics. Newton's second law denotes that the rate of change of momentum of a body is directly proportional to the force applied. Conversely, this change in momentum takes place in the direction of the applied force, which itself can depend on position (e.g., the force due to a spring). Rewriting this law, from the perspective of a single particle, in terms of a Markov blanket partition (Friston 2019), we have:

$$\dot{\mathbf{a}} = \frac{1}{m}\boldsymbol{\mu}$$
$$\dot{\boldsymbol{\mu}} = F(\mathbf{s}, \mathbf{a}) \qquad (4)$$

For a single particle, **a** and **μ** are each 3-dimensional (each spatial dimension), while **s** can be many-dimensional, as each particle it describes will have three degrees of freedom. The second law of motion is consistent with neural mechanics in terms of dynamical functions described here in the sense that they both exhibit asymmetrical flow dependencies. This ubiquitous asymmetry is the key to moving to larger spatial scales, and networks of neurons in section 4. This rests upon the structure in Figure 5, which shows the asymmetric connectivity structure between cortical columns. The neurons, which each include conductance and potential variables, now themselves become parts of sensory, active, internal, or external states with respect to a cortical column. The asymmetry now manifests in forward and backward connections along cortical hierarchies.

## 4. Cortical columns and networks

This section deals with how the same Markov blanketed structure is recapitulated at a larger spatial scale: the cortical microcircuit. Neurons are themselves components of complex self-organising systems. A key characteristic of such complex systems is that they are greater than the sum of their parts: summing up all the interactions between constituent components would not give us the full story. The properties of a complex system cannot be sufficiently understood from the level of individual components. In the present context, the brain cannot be sufficiently understood from the perspective of interactions between individual neurons. Here we appeal to the canonical microcircuit model which uses the dynamics of Figure 3 but connects the neural populations as in Figure 5. In brief, this divides neural populations into superficial and deep pyramidal cells (which turn out to be out blanket states), spiny stellate cells and inhibitory interneurons.

Considering the canonical microcircuit model has several advantages. First among these is the fact that it is used practically in the analysis of empirical brain data. This is because it can be used to specify models (i.e., hypotheses) of distributed responses – as measured with functional magnetic resonance imaging (fMRI) or electroencephalography (EEG) – that are physiologically grounded (Friston, Preller et al. 2019). For example, it is possible to specify architectures in terms of their forward and backward connections and experimental effects either as extrinsic (between region) or intrinsic (within-region) connectivity at a specific level. A third advantage is that it permits combining different imaging modalities in the form of multimodal Bayesian fusion (Wei, Jafarian et al. 2020).

The above highlights its importance for hypothesis testing. A number of neuropsychiatric conditions cannot be tackled in a compartmentalised manner and benefit from the segregation into functional units (cortical columns) offered by these microcircuits. A good example is the case of schizophrenia, in terms of the dysconnection hypothesis (Yang et al. 2015, Friston et al. 2016, Keher et al. 2008). The focus of the hypothesis is the functional disconnection of different brain regions, based on NMDA-hypofunction models of the disease. This has dramatic effects on both cortical neuronal and network activity. This hypothesis cannot be framed without knowing what is being disconnected from what. Similarly, questions about cognitive (e.g., attentional) function in health depend upon the same construct (Limanowski and Friston 2019). Other important questions, amenable to interrogation using the canonical microcircuit include questions about the nature of neurovascular coupling. For example, does it depend upon afferent presynaptic activity from extrinsic sources or (only) report to local activity mediated by recurrent (intrinsic) connectivity (Jafarian, Litvak et al. 2020)?

The brain organises itself in a decentralised way. A decentralised system, under complex systems and dynamic modelling theory, is a system whose lower-level components operate on local information to accomplish goals, i.e. control is distributed. The decentralised control is distributed such that each component of the system is equally responsible for contributing to the global, complex activity based on the component's interaction with other components (Deco et al. 2008; Chialvo, 2010; Zuo et al., 2010; Gliozzi and Plunkett, 2019; Hipólito and Kirchhoff, 2019). Figure 4 illustrates the distinction between centralised systems, i.e. a central controller exercising control (e.g. control fixed mechanism), and decentralised structures for patterns or behaviours as emergent properties of dynamical mechanisms.

Markov blankets allow us to delineate the microcircuitry connections by nuancing their intrinsic connections and how they may also change within the same network. Laminar specific connections underlie the notion of canonical microcircuit (Bastos et al. 2012). As seen in Figure 5 (second row), we can use the dependencies of this connectivity structure to provide a principled segregation into regions. Considering two columns – connected to one another – we see that if the internal and external states comprise the spiny stellate cells and interneurons of each column, the superficial pyramidal cells of one column act as the active states, while the deep pyramidal cells of the second become sensory states. Unpacking this in detail, the absence of spiny stellate or interneuron connections to the superficial pyramidal cells of other columns is consistent with the absence of influence of external on active states. The reciprocal influence is in place, allowing active states to change external states. Similarly, connections from deep pyramidal to interneurons and superficial pyramidal cells in other columns matches the directed influence of sensory over internal and reciprocated influence between sensory and active states respectively.

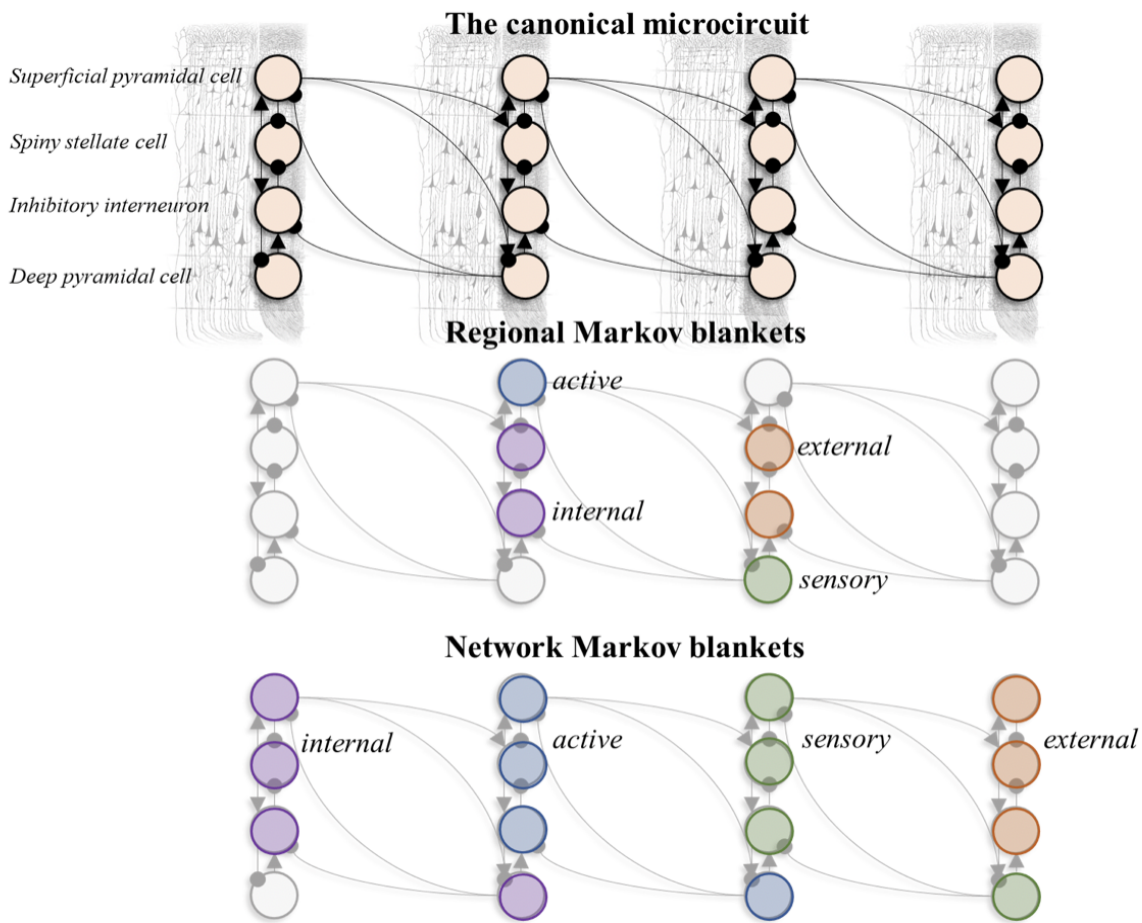

**Figure 5** *Cortical micro-circuitry*. The upper schematic shows the connectivity of the canonical microcircuit as employed for DCM (Bastos et al., 2012). This comprises four cell populations with a stereotyped pattern of connectivity. From left to right, we show forward (ascending) connections. The opposite direction shows descending connections. The dynamics of each neural population shown here obey the equations given in Figure 3, where the likelihood mappings (or **A**-matrices) in those equations specify which populations are connected to one another. As further shown by Bastos et al. (2012), feedforward connections originate predominantly from superficial layers and feedback connections from deep layers,

thus suggesting that feedforward connections use relatively high frequencies, compared to feedback connections. The second row here shows the Markov blankets that underwrite the separation into distinct cortical regions (where the superficial and deep pyramidal cells play the role of active and sensory states respectively), and the final row shows a separation into a network of regions, where the middle two regions act to insulate the far left and right regions.

What the Markov blankets in Figure 5 show is that, while a certain sparsity mediates interactions via blanket states, the internal states of a canonical microcircuit show strikingly, interconnected intrinsic architectures. In other words, we can highlight via Markov blankets, the interconnections between the neurons of origin and termination by highlighting intrinsic connectivity and extrinsic projections. This allows us to determine how top-down and bottom-up processing streams are integrated within each cortical column. Ultimately, this emphasises that intrinsic (local) behaviour is highly dependent upon extrinsic (global) behaviour via specific pyramidal populations. In short, organised patterns are observed as resulting from localised interactions without centralised control. This observation is recapitulated when we zoom out further.

Zooming out to a larger spatial scale, neuronal structures can be viewed as higher-order neural packets (Yufik and Friston 2017); i.e., as functional, larger-scale assemblies of neural packets, wrapped in their own superordinate Markov blankets. This is illustrated in the final row of Figure 5, where cortical columns now become the functional units comprising the states of a Markovian partition to define a network. Figure 6 takes this one step further, and expresses brain-wide networks as active, sensory, internal, and external states. Bounded assemblies at larger spatial scales are formed spontaneously, consistent with the self-organisation of complex systems defined as structures that maintain their integrity under changing conditions. Especially in approaches such as the one we suggest here, where coordination, segregation and integration are crucial for the self-organisation of the brain as a complex dynamic system.

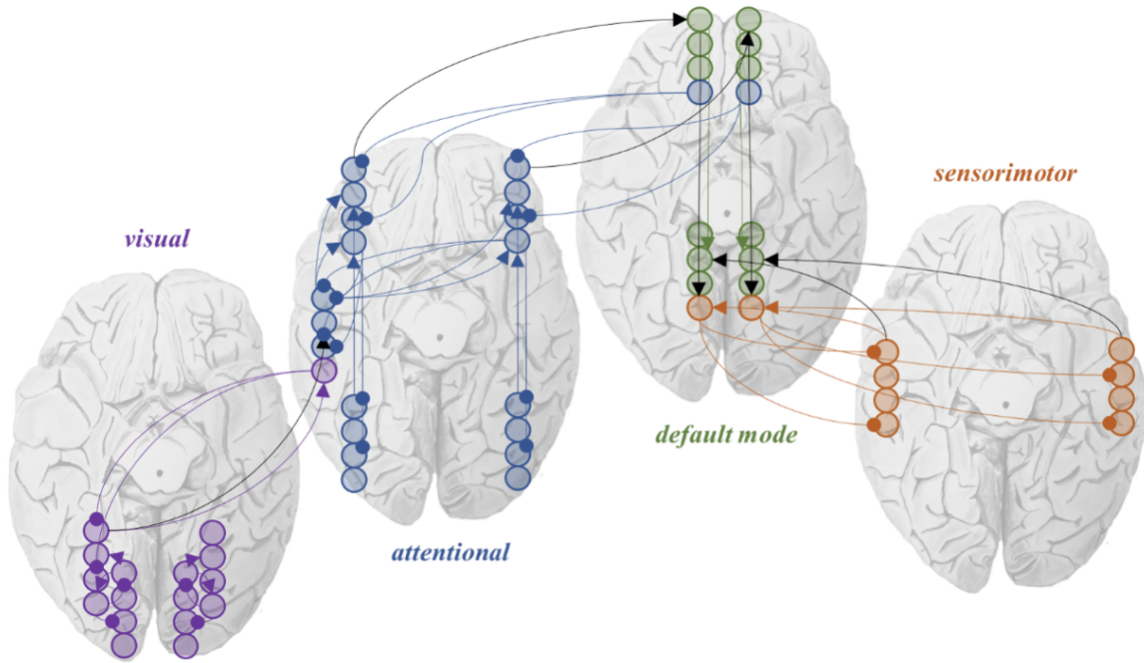

**Figure 6** *A Markov blanket of networks*. The image in this figure takes the ideas from Figure 5 one step further and shows how we could treat the connections between nodes in different networks as dependencies between states in a Markov blanketed system. Here, the networks themselves become the active, sensory, internal, and external states. This graphic is loosely structured around the kinds of networks identified using resting-state fMRI (Razi et al., 2015; Sharaev et al., 2016; Betzel et al., 2014). However, the specific connections and anatomy shown here should not be taken too seriously. Here we treat the visual networks as internal states that reciprocally influence active states (dorsal and ventral attention networks). The default mode network then plays the role of the sensory states, which mediate the influence between the above and external (sensorimotor network) states. The assignment of these is equally valid if reversed, such that sensorimotor networks become internal and visual external.

Markov blankets allow us to make salient the neuronal assemblies, as flexible but also stable biophysical structures. Put otherwise, structures such as these maintain their integrity under changing conditions. In this treatment, Markov blankets highlight the assemblies conserved over multiple levels of description, i.e. they are scale-free. Monitoring the variations in such larger spatial scales enables attributing to neurons, microcircuits, and networks the ability to undergo changes without loss of self-identity.

## 4. Discussion

The crucial point for the Markov blanket, at any scale, is that its boundaries are dictated by flows that depend upon certain states. It is by their flexibility that Markov blankets allow us to explain dynamic couplings while still drawing statistical boundaries. Markov blankets demarcate boundaries of couplings from pairs of neurons, to cortical columns and brain-wide networks. The description of neural connectivity with Markovian formalisms allows zooming in and out, identifying different functional units at different scales.

This has three practical consequences. The first is that it provides a conceptual endorsement of the DCM framework, which depends upon assessment of the effective connectivity between functionally segregated neural circuits. In brief, DCM rests upon two components: biophysical modelling using differential equations and Bayesian statistical methods for model inversion (parameter estimation) and comparison. This has many practical applications in analysing brain data acquired under a range of paradigms. For example, it has been used in the study of attentional modulation during visual motion processing (Büchel and Friston, 1997; Friston et al., 2003), in multisensory integration (Limanowski and Friston 2019), and in studies of clinical conditions (Dietz, Friston et al. 2014). Its role is to disambiguate between different hypotheses about how experimental conditions (like attentional set) modulate neuronal connectivity. With non-linear dynamic causal models (Stephan et al. 2008), non-linear DCM for fMRI enables the modelling of how activity in one population gains connection strengths, among others.

The Markovian formalism provides flexible tools to accommodate co-existing and interacting elements, which play important roles for the optimal functioning of the system. It enables us to look at the organism by considering each and every level of complexity, without losing the unity of the simplest component. Neurobiology spans from the small scale of molecular biology to the social and environmental aspects of pathology; how to accommodate these different aspects in an inter-scale manner is the key challenge, and what we are proposing is a promising tool to face it.

The persistence of Markov blanketed structures over time has a further interesting consequence. Such systems may be shown to behave according to a Bayesian mechanics (Friston 2019a) in which internal state dynamics may (on average) be expressed as gradient flows on Bayesian model evidence – or a bound on this quantity known as variational free energy. Once Markov blankets have been drawn, the neurons, cortical columns and networks, they all appear to dynamically self-organise under a common principle: the free-energy principle (Friston 2013). This says that any self-organising system will selectively interact with its environment to minimise free energy, thereby resisting the natural tendency to disorder and entropy. This paper has sought to identify the Markov blankets in the brain. We hope in future work to unpack these in terms of active inferential processes, where post-synaptic ion channels may be seen as inferring pre-synaptic channels, stellate cells and interneurons infer their counterparts in other cortical columns, and groups of columns in a network inferring the internal states of other networks.

The treatment of neurons as if they were active agents, drawing inferences about their environments, has precedence in existing theoretical work. For example, Kiebel and Friston (2011) demonstrated how dendrites can self-organise to minimise a variational free-energy bound on surprise of their presynaptic inputs, demonstrating that postsynaptic gain is itself optimised with respect to variational free-energy. This provides a principled account of neuronal self-organisation built upon the optimisation of elemental neuronal (dendritic) processing. This agenda has subsequently been developed in theoretical (Palacios et al., 2019) and empirical (Isomura and Friston 2018) studies of neuronal self-organisation.

Anticipatory mechanisms are shared by all living systems. Indeed, for an organism to remain alive, it must regulate – and therefore anticipate – the structure of its exchanges with its embedding environment, which evinces a role for *prediction*. In some organisms, especially those animals that possess a nervous system, anticipatory mechanisms are evident in patterns of organised behaviour and are made particularly evident by whole-brain dynamics over longer timescales. This motivates a specific research agenda in computational neuroscience: to investigate how microcircuits organise (and why they reorganise) on the local level and smaller, micro scales, crucially, without losing sight of the embodied brain.

It is important to recognise the limitations of this paper. While we have outlined how dynamic Markov blankets may be identified, we have done so with known equations of motion. When these are not known, as in most practical settings, the interactions between variables must be estimated. In addition, we have largely restricted our conceptual analysis to how we partition systems into fundamental (at a given scale) units. The next steps will be to unpack the consequences of this partition both analytically and through numerical simulation, with a view to the variational inferential perspective touched upon in the discussion. We have provided the foundation for this, as once we know the external and blanket states, we know what the internal states must be 'inferring'. This offers a well-formed scientific question as to the form of the implicit model the internal states use to engage in active inference – i.e., how do external states give rise to sensory states? Part of this work will be to ask questions about how brain networks self-organise. Finally, we hope to apply these ideas to the study of neuropsychiatric conditions.

Of special interest would be to develop experimental work on the span from neurobiology to social and environmental aspects of pathology, which is still missing a unifying link.

**Conclusion**

This paper investigated the characterisations of neural systems as depending upon the presence of a boundary – or Markov blanket. That is a mediation of the interaction between what is inside and outside of a system. This treatment was based upon the canonical micro-circuitry used in empirical studies of effective connectivity, to directly connect this analysis to models used in neuropsychiatric and computational psychiatry research (Frank et al. 2016; Shaw et al. 2020). The key point is that brain function depends upon the cooperative dynamics of networks, regions, and neurons. To talk meaningfully about these units of nervous tissue, we need a principled means of partitioning these from one another. This partition is afforded by the dependency and flow structure of a dynamic Markov blanket, whose structure is recapitulated each levels of analysis. This endorses the partition of neural systems at each of these stages (e.g., into neurons, regions, networks etc.), but also highlights the limitations of 'modular' perspectives on brain function that only consider a single level of description. In short, the level of analysis we choose to adopt defines a Markov blanket that tells us the appropriate functional units we need to consider. In all cases, these can be broken down into four classes of variable: active, sensory, internal, or external. In this light, the physics of the mind is consistent with the "enactive" view (Hipolito 2018), deriving cognition from an interplay between external conditions and self-organisation in the nervous system.

---

**GLOSSARY**

**Canonical Microcircuit:** Distributed network of relatively simple elements that give rise to complexity of cognitive processing by virtue of (1) their extensive interaction with other elements; and (2) their own intrinsic rich circuits. Originally introduced by (Douglas and Martin 1991) as a functional motif of interconnected neuronal populations that is considered to be replicated over the cortical sheet.

**Complex System:** a system that is composed of many components which may interact with each other. Examples include Earth's global climate, organisms the human brain, or living cells. Their behaviour is particularly difficult to model due to the dependencies and relationships between their parts and the system with the environment.

**Decentralised system:** local interactions between components of a system establish order and coordination to achieve global goals without a central commanding influence. Interactions are formed and predicated on spatiotemporal patterns, which are created through the positive and negative feedback that interactions provide.

> **Dynamic Causal Modelling:** modelling treatment of the neural dynamics as a non-linear dynamic system. Differential equations describe the interaction of neural populations, which direct or indirectly give rise to functional neuroimaging data, particularly by parameterising over directed influences or effective connectivity, usually estimated using Bayesian methods.
>
> **Emergence:** traits of a system that are not apparent from its components in isolation, but which result from the interactions, dependencies, or relationships they form when placed together in a system. These components are impossible to predict from the smaller entities that make up the system.
>
> **Neural mass models:** models of coarse-grained activity of large populations of neurons and synapses especially useful in understanding brain rhythms and synchronisation.
>
> **Non-linearity:** Non-linearity describes systems with high dependence on initial conditions, current state, and parameter values. The differential equations of non-linear dynamical systems are non-linear in the states (and parameters; in other words, they have high order terms beyond linear coupling.
>
> **Relative entropy:** mutual information, or the uncertainty about particular states minus the uncertainty, given the external states. In other words, the information gained about one set of states, given another
>
> **Self-entropy:** entropy of particular states, i.e. of states that constitute a particle, namely autonomous and sensory states. Entropy is a measure of uncertainty, disorder or dispersion
>
> **Self-organisation:** a process of spontaneous pattern formation across time scales – from microscopic cells to macroscopic organisms – that entails the emergence of stable systemic configurations that distinguish themselves from their environments.